\documentclass[fleqn]{2017SCGE}
\newcommand\aap{A\&A}                % Astronomy and Astrophysics
\newcommand\apj{ApJ}                 % Astrophysical Journal
\newcommand\apjl{ApJ}                % Astrophysical Journal, Letters
\newcommand\araa{ARA\&A}             % Annual Review of Astronomy and Astrophysics
\newcommand\mnras{MNRAS}             % Monthly Notices of the Royal Astronomical Society
\newcommand\nat{Nature}              % Nature
\usepackage{verbatim}
\usepackage{hyperref}
\usepackage{color, soul}

\begin{document}

\ensubject{subject}

%%%%%%%%%%%%%%%%%%%%%%%%%%%%%%%%%%%%%%%%%%%%%%%%%%%%%%%
%%% Authors do not modify the information below
%%% ????????????????
%%% ??????????, ????????????{}, ???????????????????
%Letter to the Editor??Article%??????
\ArticleType{Article}%??Article
\SpecialTopic{SPECIAL TOPIC: }%???????
\Year{2023}
\Month{}
\Vol{}
\No{}
\DOI{}
\ArtNo{}
\ReceiveDate{}
\AcceptDate{}
%\OnlineDate{January 1, 2016}
%%%%%%%%%%%%%%%%%%%%%%%%%%%%%%%%%%%%%%%%%%%%%%%%%%%%%%%

%%% title: ????
%%%   \title{title}{title for citation}
\title{Scintillation Arc from FRB 20220912A}{Scintillation Arc from FRB 20220912A}

%%% Corresponding author: ???????
%%%   \author[number]{Full name}{{email@xxx.com}}
%%% General author: ???????
%%%   \author[number]{Full name}{}
\author[1]{Zi-Wei Wu}{{wuzw@bao.ac.cn}}%
\author[2]{Robert A.\ Main}{}
\author[1,3]{Wei-Wei Zhu}{{zhuww@nao.cas.cn}}
\author[4]{Bing Zhang}{}
\author[1,5]{Peng Jiang}{}
\author[1,6]{Jia-Rui Niu}{}
\author[1,6]{\\Jin-Lin Han}{}
\author[1,7]{Di Li}{}
\author[1,8,9]{Ke-Jia Lee}{}
\author[10]{Dong-Zi Li}{}
\author[11]{Yuan-Pei Yang}{}
\author[12,13]{Fa-Yin Wang}{}
\author[14,15]{Rui Luo}{}
\author[1]{\\Pei Wang}{}
\author[1]{Chen-Hui Niu}{}
\author[1]{Heng Xu}{}
\author[8,9]{Bo-Jun Wang}{}
\author[6,8]{Wei-Yang Wang}{}
\author[1,6]{Yong-Kun Zhang}{}
\author[16]{\\Yi Feng}{}
\author[1,6]{De-Jiang Zhou}{}
\author[17]{Yong-Hua Xu}{}
\author[18]{Can-Min Deng}{}
\author[1,6]{Yu-Hao Zhu}{}

%%% Author information for page head. ?¨¹?§Ö????????
%%% ??????????????, ??????????author???
\AuthorMark{Wu Z.}%\authorcr????????

%%% Authors for citation. ????????§Ö????????
%%% ??????????????, ??????????author???
\AuthorCitation{Ziwei Wu, Robert A.\ Main, Weiwei Zhu, et al}

%%% Address. ???
%%%   \address[number]{Address, City {\rm Postcode}, Country}
\address[1]{National Astronomical Observatories, Chinese Academy of Sciences, Beijing 100101, China}
\address[2]{Max-Planck-Institut f\"ur Radioastronomie, Auf dem H\"ugel 69, 53121 Bonn, Germany}
\address[3]{Institute for Frontier in Astronomy and Astrophysics, Beijing Normal University, Beijing 102206, China}
\address[4]{Nevada Center for Astrophysics and Department of Physics and Astronomy, University of Nevada, Las Vegas, NV 89154, USA}
\address[5]{CAS Key Laboratory of FAST, National Astronomical Observatories, Chinese Academy of Sciences, Beijing 100101, China}
\address[6]{School of Astronomy, University of Chinese Academy of Sciences, Beijing, China}
\address[7]{NAOC-UKZN Computational Astrophysics Centre, University of KwaZulu-Natal, Durban 4000, South Africa}
\address[8]{Department of Astronomy, Peking University, Beijing 100871, China}
\address[9]{Kavli Institute for Astronomy and Astrophysics, Peking University, Beijing 100871, China}
\address[10]{Cahill Center for Astronomy and Astrophysics, MC 249-17 California Institute of Technology, Pasadena CA 91125, USA}
\address[11]{South-Western Institute For Astronomy Research, Yunnan University, Yunnan 650504, China}
\address[12]{School of Astronomy and Space Science, Nanjing University, Nanjing 210093, China}
\address[13]{Key Laboratory of Modern Astronomy and Astrophysics (Nanjing University), Ministry of Education, China}
\address[14]{CSIRO Space and Astronomy, PO Box 76, Epping, NSW 1710, Australia}
\address[15]{Department of Astronomy, School of Physics and Materials Science, Guangzhou University, Guangzhou 510006, China}
\address[16]{Research Center for Intelligent Computing Platforms, Zhejiang Laboratory, Hangzhou 311100, China}
\address[17]{Yunnan Observatories, Chinese Academy of Sciences, Kunming 650011, China}
\address[18]{Guangxi Key Laboratory for Relativistic Astrophysics, Department of Physics, Guangxi University, Nanning 530004, China}

%\contributions{}%????????

%%% Abstract. ??
\abstract{
We present the interstellar scintillation analysis of fast radio burst (FRB) 20220912A during its extremely active episode in 2022 using data from the Five-hundred-meter Aperture Spherical Radio Telescope (FAST). 
We detect a scintillation arc in the FRB's secondary spectrum, which describes the power in terms of the scattered FRB signals' time delay and Doppler shift. 
The arc indicates that the scintillation is caused by a highly localized region.
Our analysis favors a Milky Way origin of the ionized interstellar medium (IISM) for the localized scattering medium but cannot rule out a host galaxy origin.
We present our method for detecting the scintillation arc, which can be applied generally to sources with irregularly spaced bursts or pulses.
These methods could help shed light on the complex interstellar environment surrounding the FRBs and in our Galaxy.
}

%%% Keywords. ?????
\keywords{Fast Radio Burst, Scintillation, FRB~20220912A}

\PACS{95.85.Bh, 98.38.Dq, 78.70.Ps}

\maketitle

%\tableofcontents%?????

%%%%%%%%%%%%%%%%%%%%%%%%%%%%%%%%%%%%%%%%%%%%%%%%%%%%%%%
%%% The main text. ???????
%???????????????????\cref{fig1}
%\twocolumn\onecolumn
%%%%%%%%%%%%%%%%%%%%%%%%%%%%%%%%%%%%%%%%%%%%%%%%%%%%%%%
\begin{multicols}{2}
\section{Introduction}\label{section1}
The ionized interstellar medium (IISM)'s refractive index 
%\Authorfootnote
%\noindent
fluctuations cause phase variations when rays of radio signals from compact sources pass through.  
The interference between these scattered rays modulates the signal intensity as a function of frequency and time, which is well-known as interstellar scintillation (ISS) \cite{sch68}. 
Two main types of ISS are diffractive ISS (DISS) \cite{ric69} and refractive ISS (RISS) \cite{sie82, rcb84}).
DISS is interference, resulting in significant modulations on short timescales. At the same time, RISS is caused by large-scale re-focusing and de-focusing of the rays, resulting in small modulations on long timescales \cite{ric90}.

ISS phenomena are often observed from pulsars \cite{ric90, lvm+22} and quasars \cite{wagw95, bts+22}.
Compact sources affected by strong ISS phenomena often show a highly variable dynamic spectrum, the radio flux density as a function of time and frequency. 
DISS %scintillation, 
is characterized by scintillation bandwidth ($\Delta \nu_{\rm{d}}$) and scintillation timescale ($\tau_{\rm{d}}$) in the frequency and time domains \cite{ric90}, respectively. 
In around the 2000s, scintillation arcs were discovered in pulsars, which are parabolas in the 2D power spectra which arise from the expected dependence of the Doppler rate and the time delay in a thin screen \cite{rlg97, smc+01}. 
Such phenomena opened a new window to study the IISM and pulsars \cite{wks+08, bmg+10, rcb+20,yzm+21, zty+23}.
Scintillation arcs have been detected from a number of pulsars \cite{wvm+22, srm+22, mpj+23}, suggesting that scattering in the IISM is often highly localized.

Fast Radio Bursts (FRBs) are a new type of transient radio source with pulse lengths ranging from a fraction of a millisecond to a few milliseconds, first discovered by Lorimer et al. 2007 \cite{lbm+07}.
Due to their small angular sizes, FRBs are scintillating sources as well \cite{mls+15, rsb+16, cordes+19}.
They also exhibit another multi-path propagation effect called scattering.
The primary scattering mechanism of radio signals is diffraction caused by random fluctuations of the IISM.
The pulse is consequently broadened and exhibits a quasi-exponential tail.
Simultaneous detection of ISS and scattering seems to be unrealistic at the same frequency for MW pulsars \cite{cr98, wcv+23}.
In some cases, however, they are detected simultaneously from FRBs at the same band \cite{mls+15, occ+22}, suggesting FRB scattering from two screens, one in the Milky Way (MW) and the other in the FRB host galaxy \cite{mls+15}.
Recently, the annual variation in scintillation timescale from FRB~20221024A has allowed the measurement of its screen distance, indicating that the dominant screen is in the local environment 400$\pm$40 or 460$\pm$60\,pc from Earth, potentially associated with the Local Bubble \cite{mbm+23}.
The distance and geometry of scattering screens can be more reliably derived from scintillation arcs \cite{rcb+20}, 
as the arc curvatures depend primarily on the relative distances and velocities of the source, screen, and observer, and not on the variation in scintillation strength. However, scintillation arcs have yet to be clearly detected from FRBs.

In this paper, we present the first detection and analysis of scintillation arcs from an FRB, FRB 20220912A,
a highly active FRB with dispersion measure (DM) $\sim$220 pc~cm$^{-3}$. 
We organize this paper as follows: in Section~2, we describe our observations and data processing; 
in Section~3, we show the analysis and results; and  
Section 4 contains the conclusions. 

\begin{figure}[H]
    \centering
    \includegraphics[width=0.99\linewidth]{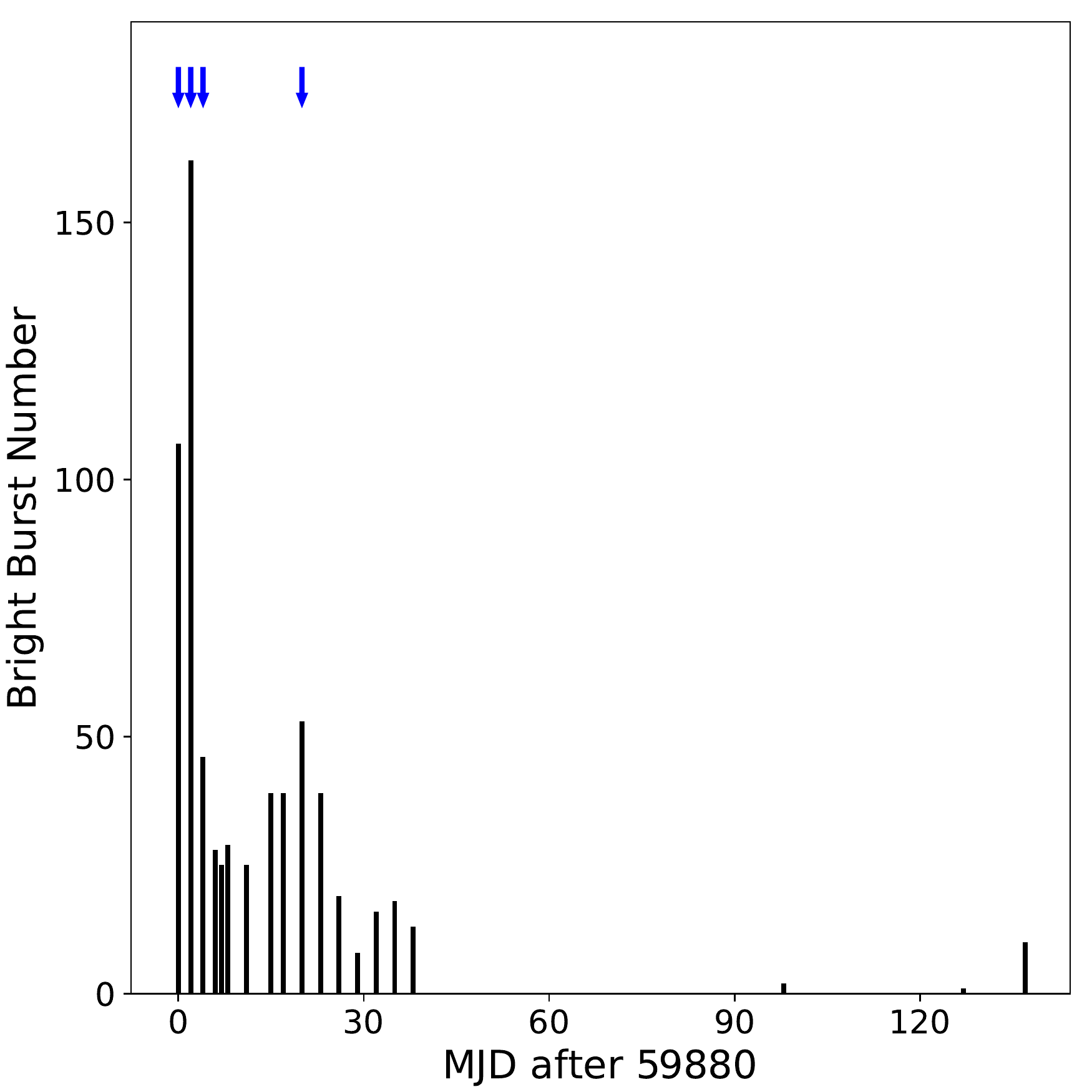}
    \caption{FRB~20220912A bright burst (S/N $\geq$ 10) detection number. 
    The blue arrows represent the time with clear scintillation arc detection.}
    \label{fig:burst_event_rate}
\end{figure}

\section{Observations and Data Processing} \label{sec:observations and data processing}

FRB 20220912A is a repeating FRB detected by various telescopes during an active episode from October 2022 to early 2023 \cite{mc+22, znf+22, ydn+22}.
Here we report the scintillation properties of FRB 20220912A, observed from 2022 October 28 to 2023 February 14 (see Figure~\ref{fig:burst_event_rate}), with the central beam of the FAST 19-beam L-band array \cite{jyg+19, jth+20, qys+20}, pointing to R.A. = 23$^{\rm{h}}$09$^{\rm{m}}$04.9$^{\rm{s}}$, Dec = +48$^{\circ}$42$^{'}$25.4$^{''}$ \cite{rav22}. 
The corresponding galactic longitude and galactic latitude in deg are 
106.5 and -10.7, respectively.
The FAST search mode data processing applied in this work is described in detail by Niu et al.2022 \cite{nzz+22}, including de-dispersion and searching the data using \textsc{presto}\footnote{\url{https://github.com/scottransom/presto}} \cite{ran11}.
In the end, the recorded data are in the \textsc{psrfits} format with four polarizations,
49.152 $\mu$s sampling interval and 4096 frequency channels.
The burst detection and analysis of burst properties are presented in a companion paper \cite{zlz+23}.

\begin{figure}[H]
\centering
\includegraphics[scale=0.4]{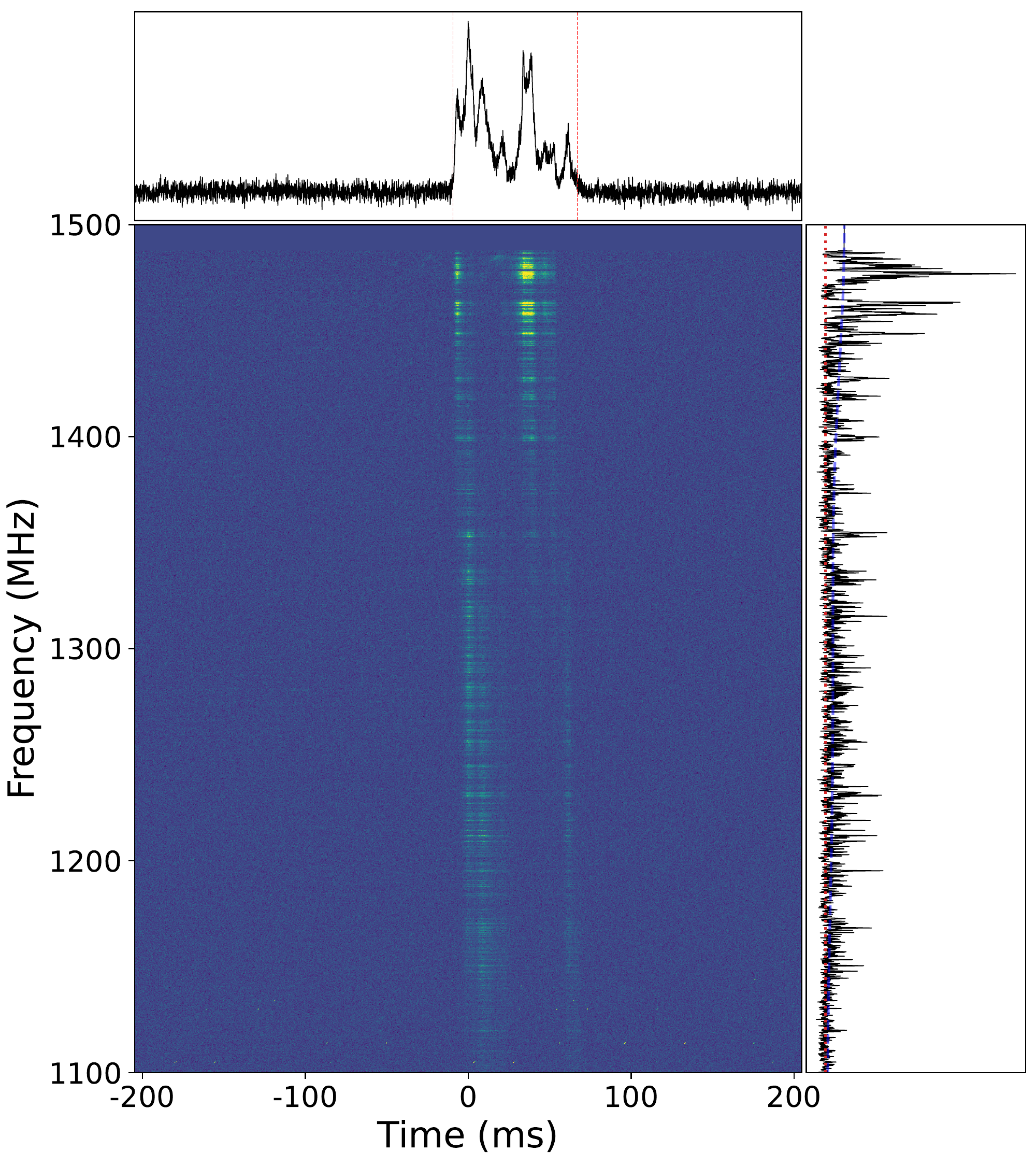}
\caption{DM corrected (220 pc~cm$^{-3}$) time-frequency data of one burst detected by FAST on October 28, 2022, from FRB~20220912A. In the top sub-panel, the frequency-averaged profile is black, while the red dotted lines enclose the ``on'' window of each FRB. The intensity distributions in the frequency domain are shown in black in the right sub-panel and are smoothed with a Gaussian function with a width of 1/8 of the whole frequency band in blue. The red dotted line in the right sub-panel is at the rms of the background.} 
\label{fig:dynamic}
\end{figure}

\begin{figure}[H]
    \centering    
    \includegraphics[width=0.99\linewidth]{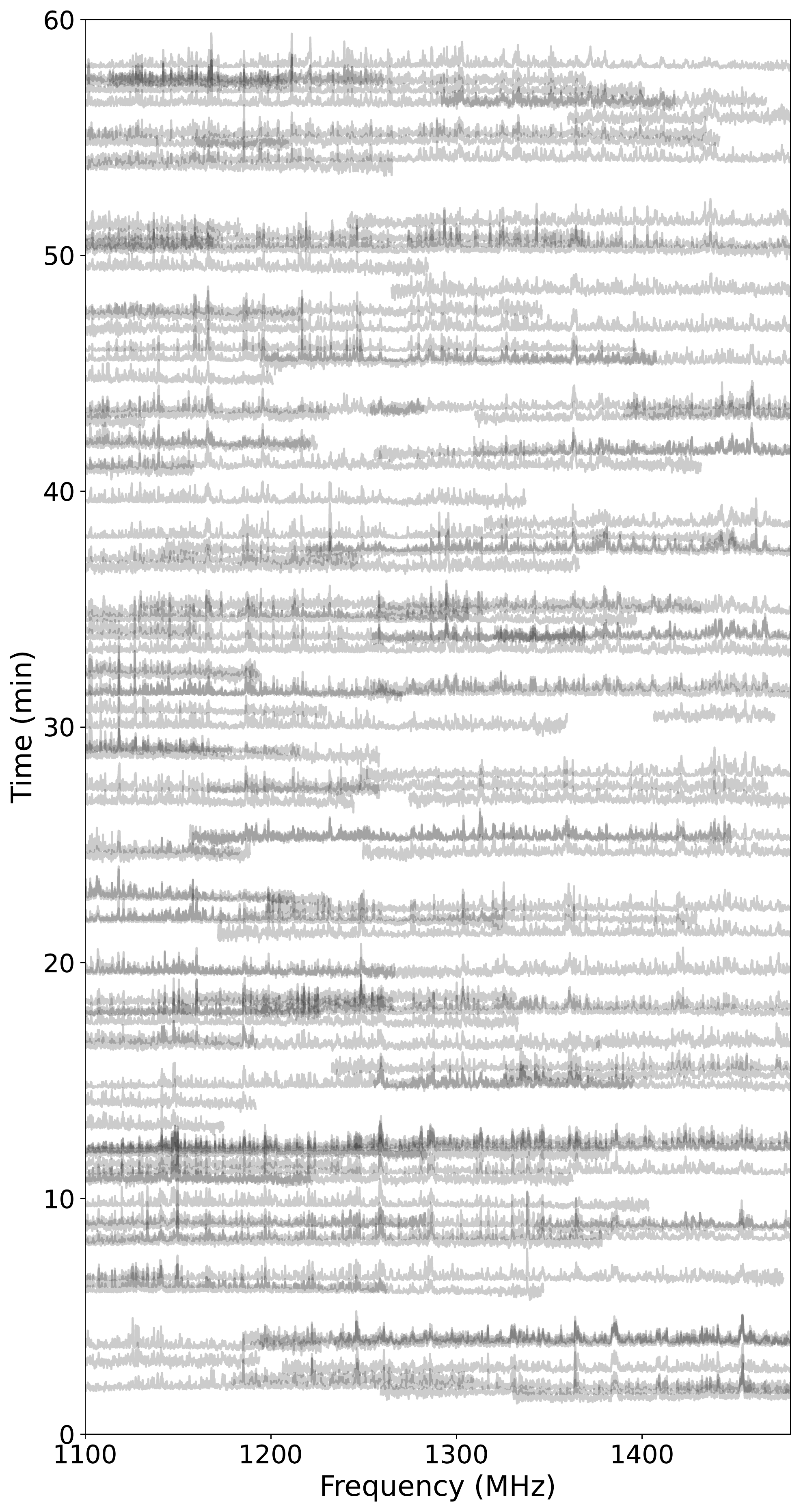} \\
    \caption{Burst spectra versus time (dynamic spectrum) for FRB~20220912A on MJD 59882 with FAST. The frequency range of the burst spectrum is restricted to the sub-channels where the burst is significantly detected. Nearby burst spectra are clearly highly correlated, indicating that these bursts are imprinted with the same scintillation pattern, which also allows a robust scintillation timescale measurement. The same printed contrast of each burst is applied. The lines have a transparency.}
    \label{fig:data_panorama}
\end{figure}

\begin{figure}[H]
    \centering
    \includegraphics[width=0.96\linewidth]{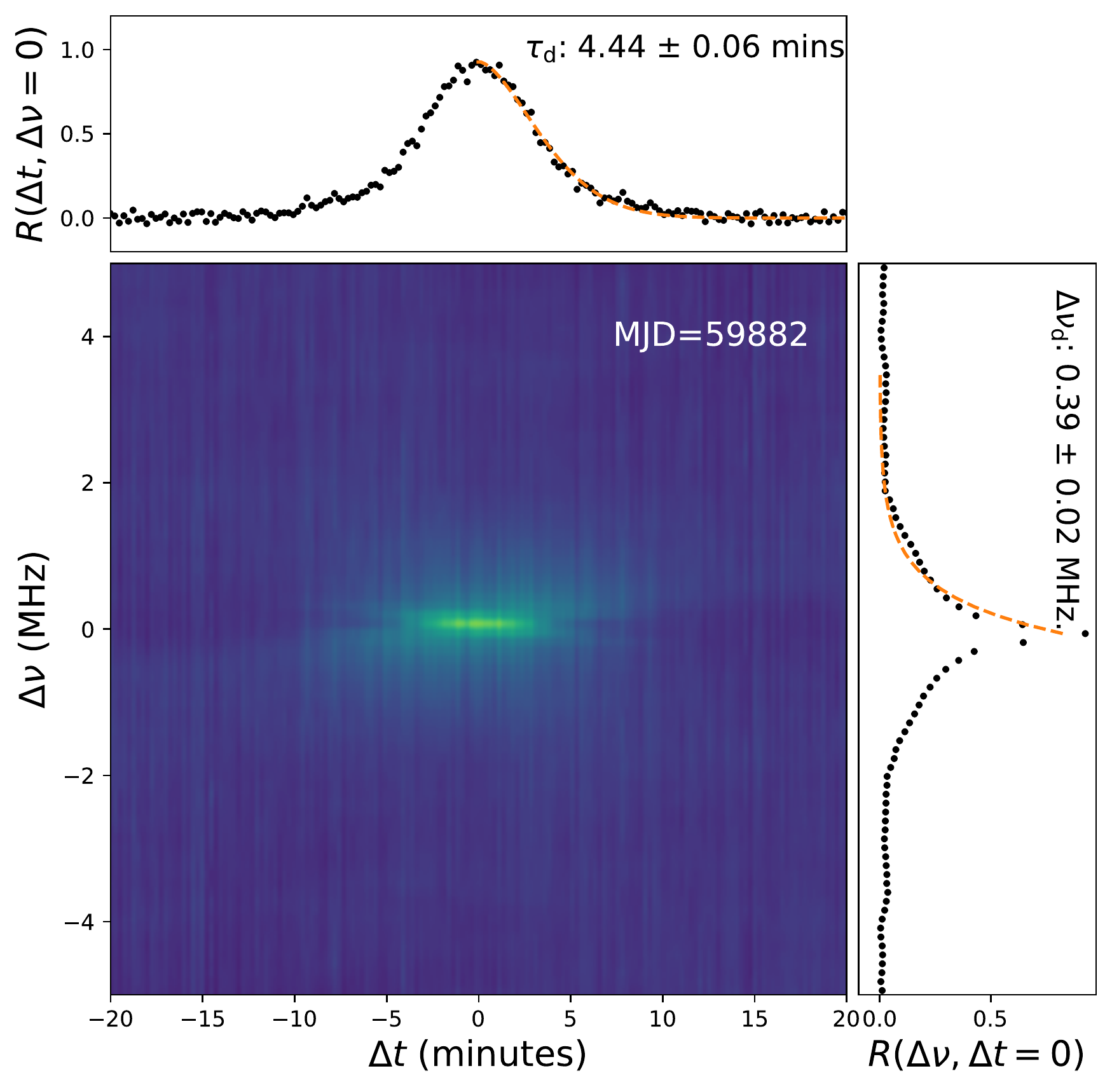}
    \caption{The 2D ACF plot of FRB~20220912A on MJD~59882 with FAST. In the two smaller side plots, the black points are the 1D ACFs at zero frequency and time lag, the orange curves are the best fits from which the scintillation timescale $\tau_{\rm{d}}$ and scintillation bandwidth $\Delta \nu_{\rm{d}}$ are derived, respectively. The color scale of the 2D ACF is in the range of -0.2 to 1.2.}
    \label{fig:2dacf}
\end{figure}

Before extracting the burst spectra, radio-frequency interference is identified and then masked. 
For each burst, we get a spectrum by averaging in the ``on'' window, dividing by its smoothed spectrum, keeping only the frequency band where the burst power is larger than the rms of the background (see Figure \ref{fig:dynamic}), following Main et al.2022b \cite{mhm+22}.
Afterwards, we have N burst spectra $F(t_{\rm{i}}, \nu)$ for each observation (see Figure~\ref{fig:data_panorama}).
The bursts used for this scintillation study have a signal-to-noise ratio (S/N\footnote{The S/N is given by the ratio of the maximum intensity of pulse profile to the standard deviation of the off-pulse region.}) greater than 10.
We note that the detection threshold of S/N is selected to be 7 in other FAST FRB works \cite{zhz+22}. 
We use a higher threshold as studies of scintillation arcs require sensitivity to small fluctuations, often from deflected ray paths with $\lesssim$1\% of the FRB's mean flux.
For each burst, we calculate the spectrum's frequency autocorrelation function (ACF).
We then construct the 2D ACFs (see Figure~\ref{fig:2dacf}) by binning the frequency ACFs of all burst pairs as a function of time separation $\Delta t$ in 0.25 min bins, weighted by the errors derived in the same manner as Main et al.2022b \cite{mhm+22}.

A Gaussian function is fit to all values of a slice of the 2D ACF$(\Delta t, \Delta \nu = 0)$, and the scintillation timescale $\tau_{\rm{d}}$ is the width at 1/e height. 
The scintillation bandwidth is typically defined as the half-width at the half maximum of a slice 2D ACF$(\Delta t = 0, \Delta \nu)$ with an exponential function, similar to Reardon et al. 2019 \cite{rch+19}.
With a Hanning window function applied to the whole ACF, we perform the 2D Fourier transform of the ACF to compute the secondary spectrum \cite{rlg97,smc+01}, shifting it and then converting the relative power levels into a decibel scale.

\begin{figure*}
\centering
\includegraphics[width=0.49\linewidth]{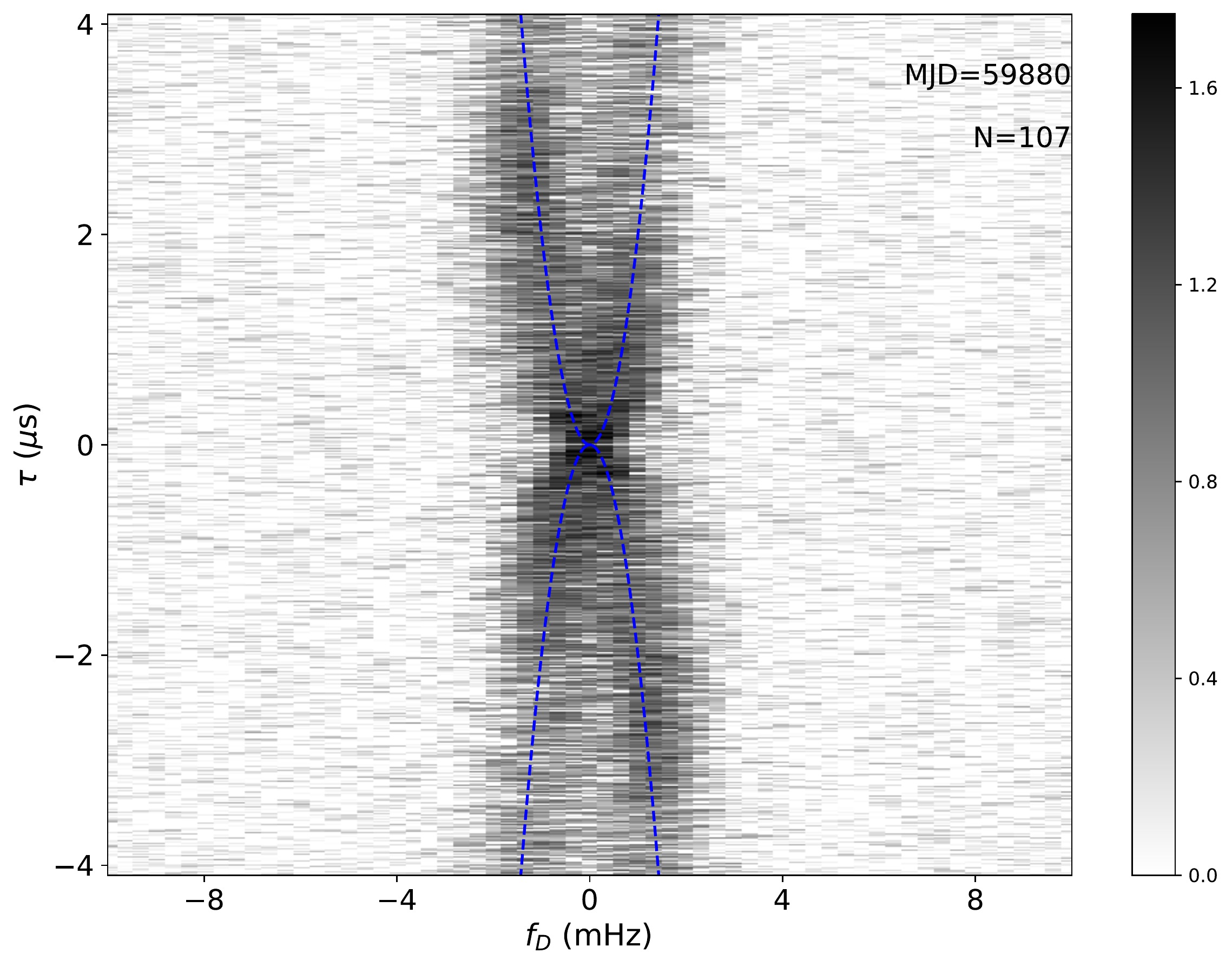} 
\includegraphics[width=0.49\linewidth]{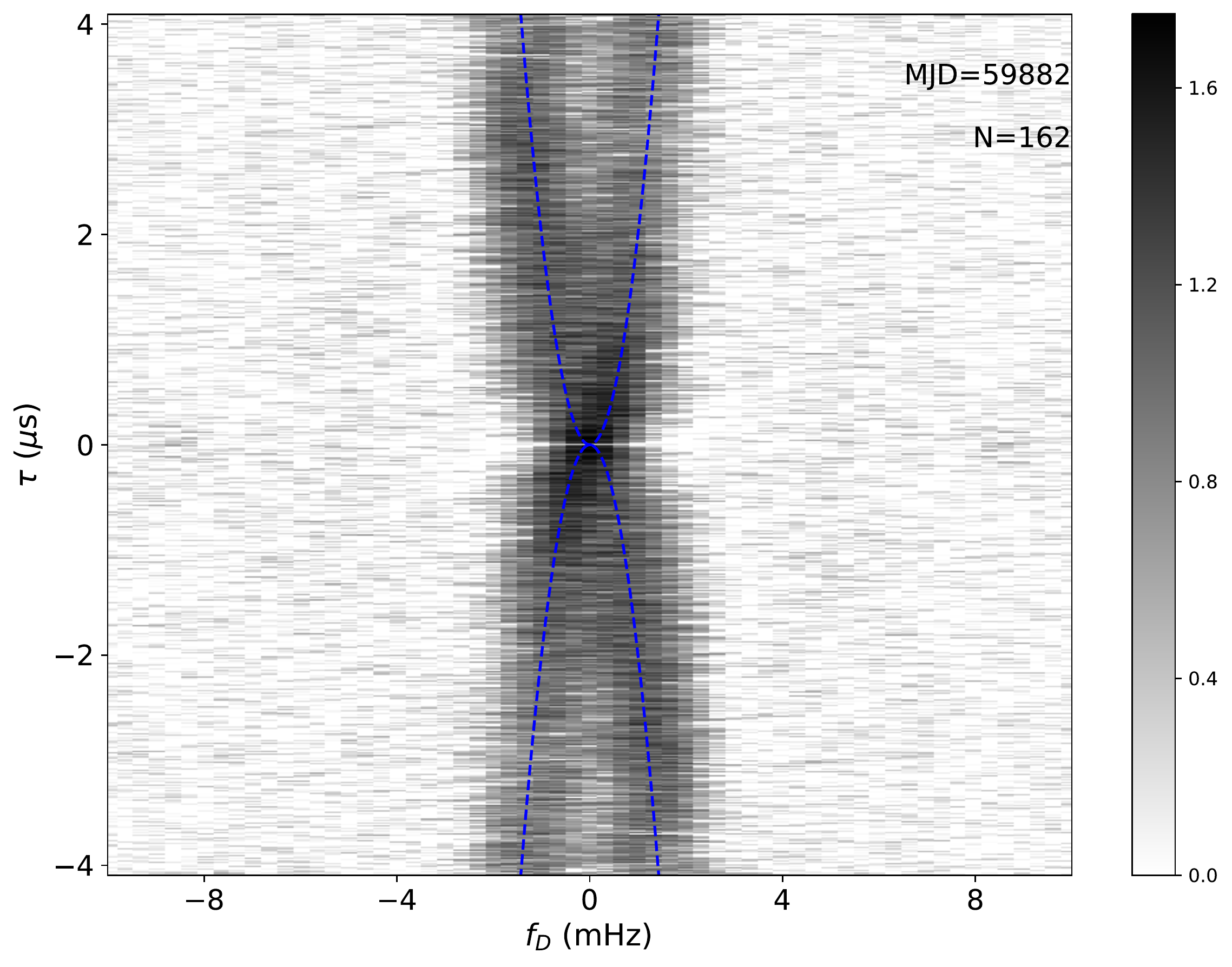} \\
\includegraphics[width=0.49\linewidth]{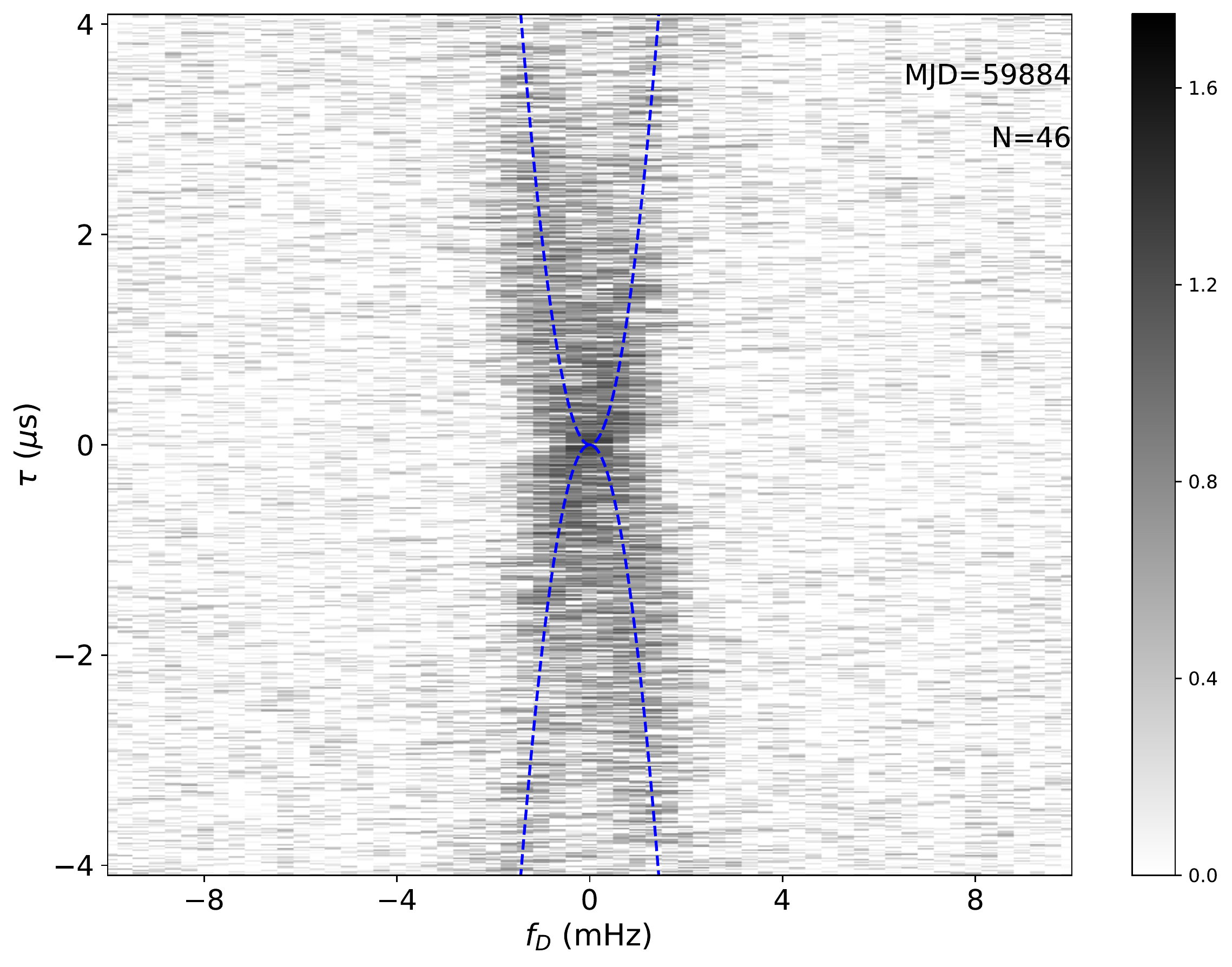} 
\includegraphics[width=0.49\linewidth]{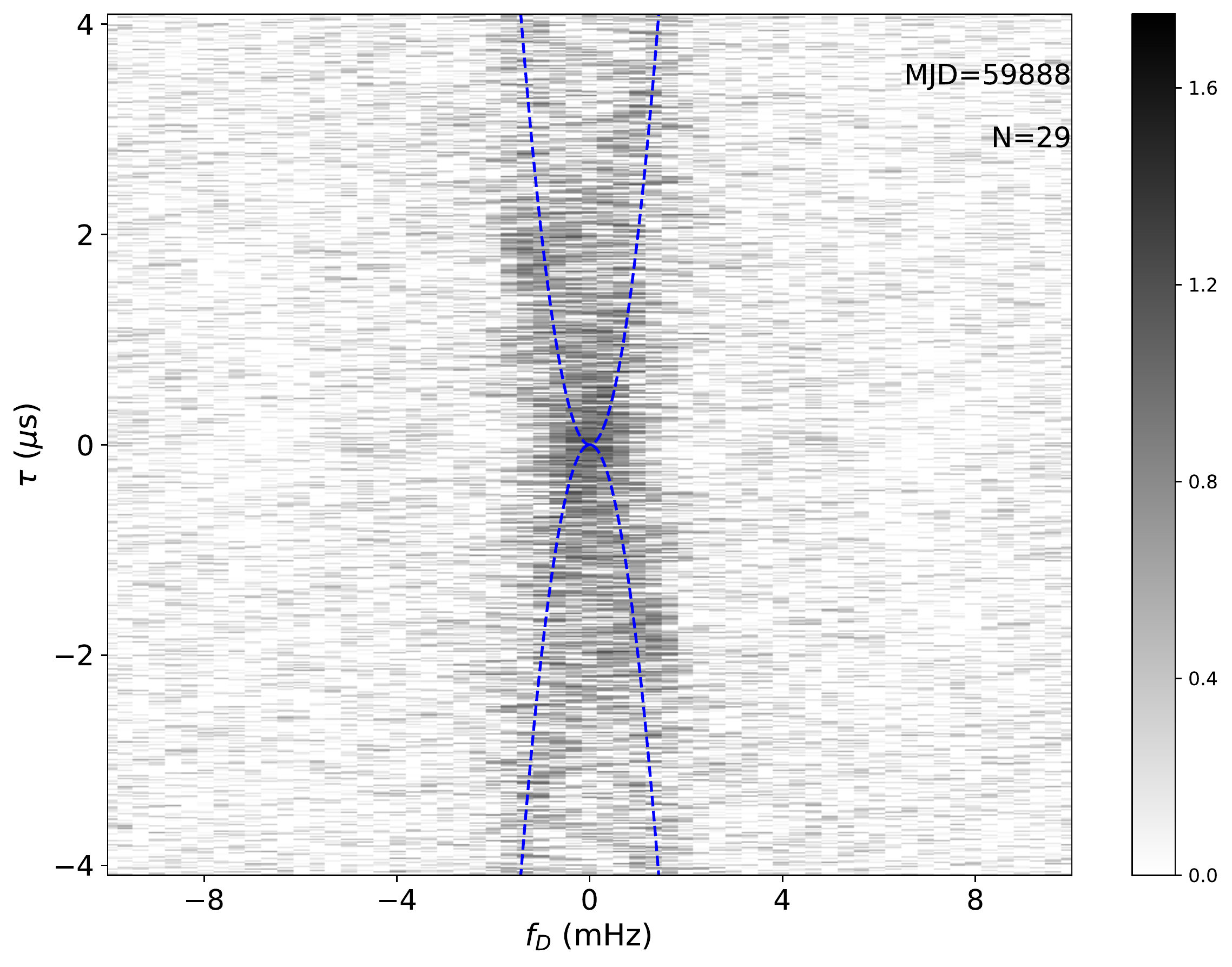} \\
\includegraphics[width=0.49\linewidth]{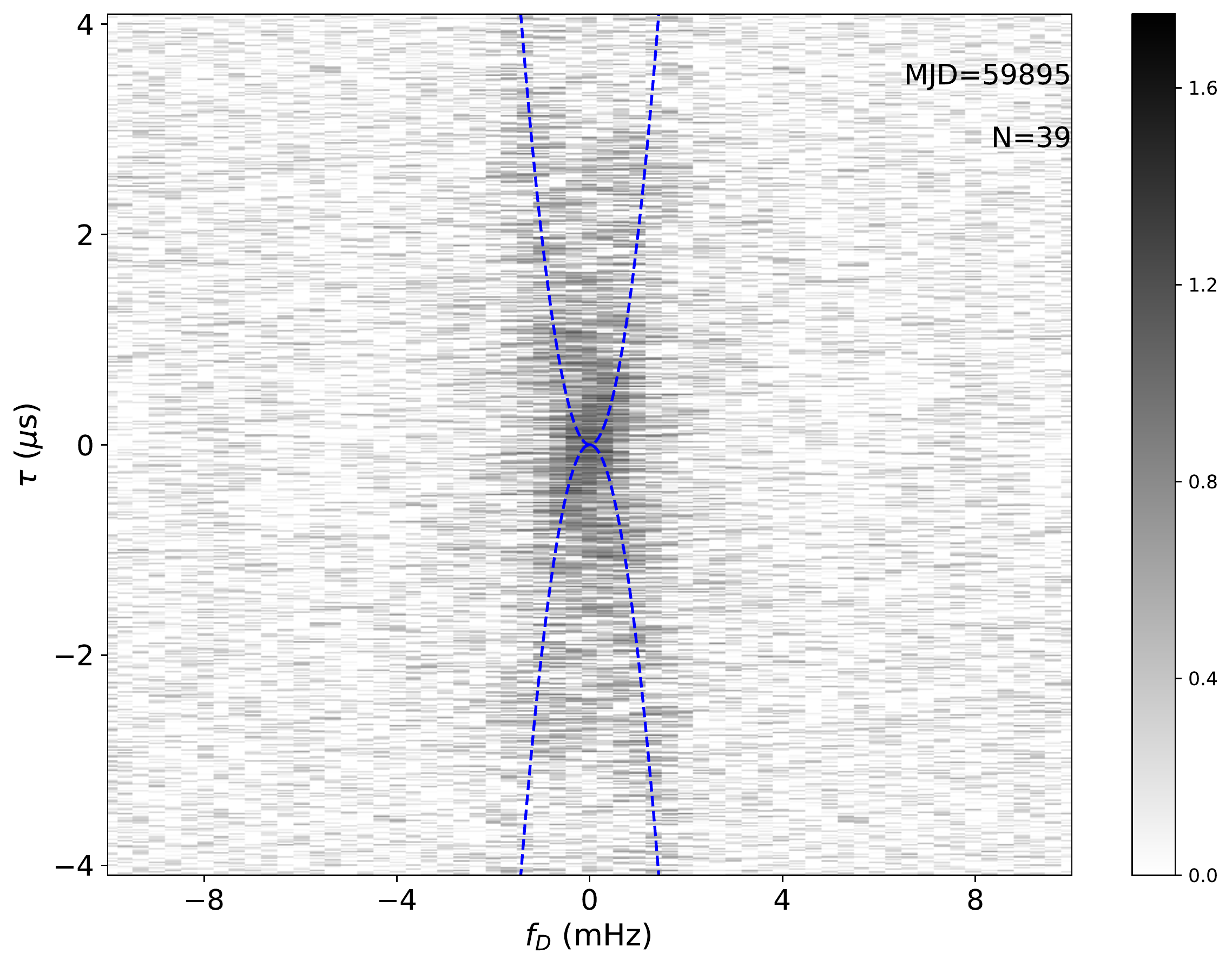} 
\includegraphics[width=0.49\linewidth]{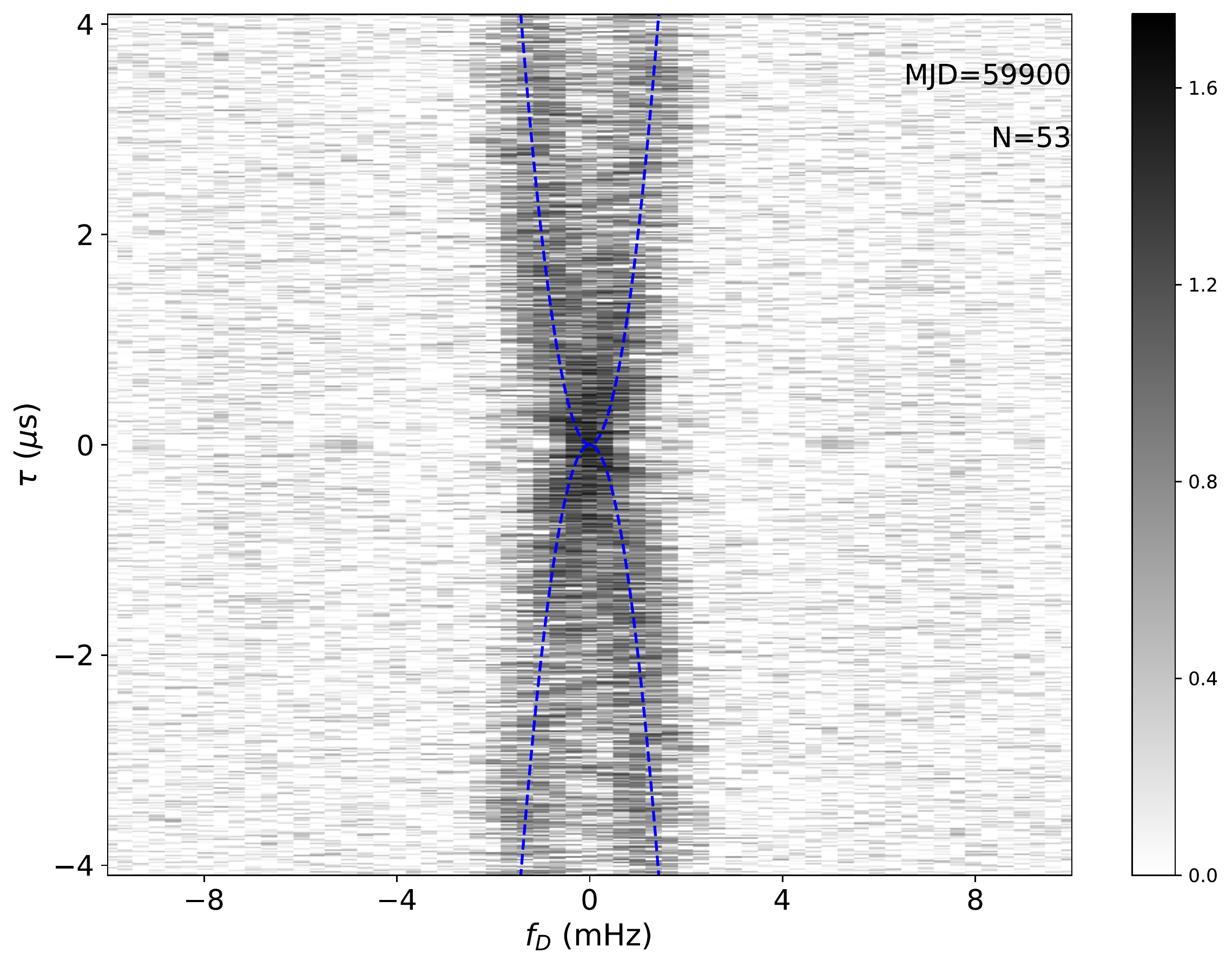} \\
\caption{The secondary spectra of FRB~20220912A from the epochs with bright (S/N $\geq$ 10) burst detection rate greater than 50 counts per hour. The overlaid parabolas in blue correspond to an arc curvature $\eta$ of 1.97$\pm$0.06\,$s^{3}$. The color scale is logarithmic.}
\label{fig:arc_zoo}
\end{figure*}

\section{Analysis and Results}
We study the 2D ACF and the 2D power spectrum of scintillation, often known as the `secondary spectrum', which is the 2D Fourier transform of the ACF \cite{smc+01}.  
The secondary spectrum describes the power as a function of Doppler shift ($f_{\rm{D}}$) and geometric time delay ($\tau$) of each scattered ray with respect to the line of sight \cite{wms+04,crs+06}. 
The interference between these rays often shows a parabolic arc $\tau = \eta f_{\rm{D}}^{2}$ owing to the shared dependence of $\tau$ and $f_{\rm D}$ on $\theta$, with the curvature $\eta$ depending on the relative distance and geometry of the source, screen, and observer.
In the following, we ignore the contribution from the host galaxy redshift on scintillation since it is only 0.0771 \cite{rcc+22}.

\subsection{Scintillation Timescale}
The scintillation timescale ($\tau_{\rm{d}}$) is determined by the scattering material properties and the effective velocity of the entire scintillation system.
The difficulty of measuring $\tau_{\rm{d}}$ of an FRB is that many burst pairs need to be detected within $\tau_{\rm{d}}$,
which has been only possible in a small number of FRBs to date. 
The measurable scintillation timescale $\tau_{\rm{d}}$ depends on the bright burst detection rate and the broadband frequency structure. 

With FAST observations, the burst spectra of FRB~20220912A for all selected bursts detected on 2022 October 30 are exhibited in Figure~\ref{fig:data_panorama}.
Nearby bursts correlate at $\sim100\%$, with a decaying trend in correlation coefficient over time lag in Figure~\ref{fig:2dacf}.
The resulting scintillation timescale $\tau_{\rm{d}}$ and scintillation bandwidth $\Delta \nu_{\rm{d}}$ of FRB~20220912A at Modified Julian Day (MJD) 59880, which is 4.44$\pm$0.06 minutes and 0.39$\pm$0.02~MHz, respectively. 
We note that since the ``on'' window determination for each burst depends on the threshold limit, the processing becomes ``non-linear''. 
This makes quantitative error analysis and parameter estimation subject to the threshold limit chosen.
At lower observing frequency for all observations, the values of scintillation bandwidth $\Delta \nu_{\rm{d}}$ are comparable to or even lower than the frequency resolution.
Consequently, the measurements of the scintillation bandwidth $\Delta \nu_{\rm{d}}$ are noisy and heavily biased, resulting in overestimated scintillation bandwidth $\Delta \nu_{\rm{d}}$. 
Thus, the screen location cannot be constrained from scintillation bandwidth \cite{cr98, rch+19}.

\subsection{Detecting Scintillation Arcs}
We present the secondary spectra of FRB~20220912A for the epochs with the bright burst rate $\geq$ 50 counts per hour in Figure~\ref{fig:arc_zoo}.
%For the first time from an FRB, scintillation arcs are detected.
The scintillation arc detected from the secondary spectra indicates a thin scattering screen dominating the scintillation of FRB~20220912A.
The most significant arc was observed on MJD~59882, with the most burst samples on one day (see Figure~\ref{fig:data_panorama}).
Scintillation arcs are also detected in other observations taken on MJD~59880, MJD~59884, and MJD~59900, respectively.
%The Doppler shift of all the scattered waves for these arcs is distributed within $\leq$ 2~mHz, and the significant power extends to the maximum time delay of 4.0~$\mu$s from some observations.
The secondary spectrum usually shows a power asymmetry, which indicates a phase gradient across the dominant scattering screen \cite{crg+10}. 
The scintillation arcs become weaker in later observations, likely caused by the lower event rate  ( Figure~\ref{fig:burst_event_rate}). 
The measured scintillation arc curvature on MJD~59882 is 1.97$\pm$0.06~s$^{3}$ with \textsc{parabfit} described in Bhat et al.2016 \cite{bot+16} based on a Hough transform (see Figure.~\ref{fig:hough}).
The corresponding scaled scintillation velocity $W_{\eta}$ can be obtained from $\lambda/\sqrt{2c\eta}$ \cite{mpj+23}, which is 38.5$\pm$0.6 km s$^{-1}$ $\rm{kpc}^{-1/2}$.
Then we overlay this arc curvature to all the secondary spectra (see Figure \ref{fig:arc_zoo}). 
We do not find any significant change in the scintillation arc curvature among these arc detections.
%The absence of the scintillation arc in other epochs likely results from the small number of bright bursts detected. 

\subsection{Scintillation in host galaxy or MW}
IISM in the MW or the FRB host galaxy can both cause scintillation.
In this section, we aim to determine the origin of the scintillation arcs.
For a thin screen, the arc curvature $\eta$ can be derived as \cite{crs+06}:
\begin{equation}
 \eta = \frac{D_{\rm{FRB}}s(1-s)}{2\nu^{2}} \frac{c}{(\boldsymbol{V_{\rm{eff}}} \cos\boldsymbol{\psi})^{2}}.
 \label{arc_curature}
\end{equation}
where $c$ is the speed of light; $\nu$ is the observing frequency; $D_{\rm{FRB}}$ is the distance between the FRB and the observer; $s$ is the fractional distance of the scattering screen ($s = 1 - d_{\rm{sc,MW}}/D_{\rm{FRB}}$ in which $d_{\rm{sc,MW}}$ is the screen distance from Earth in MW);
and $\boldsymbol{\psi}$ is the angle between the (1D) scattering structure and $\boldsymbol{V_{\rm eff}}$, and $\boldsymbol{V_{\rm eff}}$ is the effective scintillation velocity vector at the scattering screen. 
Depending on the relative velocities of the FRB $\boldsymbol{V_{\rm{FRB}}}$, the possible FRB orbital motion $\boldsymbol{V_{\rm o}}$, Earth motion $\boldsymbol{V_{\rm E}}$ and motion of the screen $\boldsymbol{V_{\rm sc}}$, in the form of \cite{cr98}:
\begin{equation}
      \boldsymbol{V_{\rm{eff}}} = (1-s) (\boldsymbol{V_{\rm{FRB}}} + \boldsymbol{V_{\rm{o}}}) + s \boldsymbol{V_{\rm{E}}} - \boldsymbol{V_{\rm{sc}}(s)}.
          \label{eq:veff}
\end{equation}

%\textcolor{blue}{In the following contents, we assume $\cos(\psi)=1$, indicating $\boldsymbol{V_{\rm eff}}$ is aligned with the screen, or the screen is isotropic.}

\subsubsection{Scintillation in host galaxy}
Considering now, for distant FRBs and scintillation effects dominated by IISM in their surrounding environment ($s \rightarrow$ 0), Eq.~\ref{arc_curature} can be simplified as:
\begin{equation}
\begin{split}
\eta \approx \frac{cd_{\rm{sc,HG}}}{2\nu^{2}} \frac{1}{[(\boldsymbol{V_{\rm{FRB}}} + \boldsymbol{V_{\rm{o}}} - \boldsymbol{V_{\rm{sc}}}) \cos\boldsymbol{\psi}]^{2}}, %\,\,\, \mathrm{s} \sim 0 \\
\label{eq:arc_definition_hg}
\end{split}
\end{equation}
where $d_{\rm{sc, HG}}$ is the distance between the screen and the FRB in the host galaxy.

Let us assume that a stationary and isotropic screen caused the arc in the host galaxy.
One of the more prominent models for the origin of FRBs is a neutron star, possibly young or highly magnetized \cite{msh+18}.
This model is reinforced by the luminous radio bust FRB~200428 event from Galactic magnetar SGR 1935+2154\cite{brb+20}. 
Young pulsars are often born with high birth-kick velocity \cite{ll94}.
If we assume that FRB~20220912A moves with a high transverse velocity of $\sim$500~km s$^{-1}$, the derived distance $d_{\rm{sc, HG}}$ from the FRB to the dominated screen in the host galaxy is $\sim$170~kpc. 
Alternatively, a young neutron star is sometimes surrounded by a shell of a supernova remnant.
The remnant is often located at a distance of $\sim$10~pc from the pulsar. 
In this case, a small FRB velocity of $\sim4\,\mathrm{km}\, \mathrm{s}^{-1}$ is required from equation \ref{eq:arc_definition_hg} if the scintillation arc is caused by the supernova remnant shell.  
Finally, if an FRB is produced by a Crab-like pulsar with a transverse velocity of 100~km s$^{-1}$ \cite{alp75} and the detected scintillation arc is caused by its surrounding environment within 2~pc \cite{lpg01}, we expect an arc curvature of $\sim$6$\times10^{-4}$~s$^{3}$ at 1.25~GHz.
Similar arc curvature has been detected from PSR J0538+2817 associated supernova remnant S147 \cite{yzm+21}.
Thus, searching for small arc curvature presents a valuable opportunity to investigate key characteristics of FRBs surrounding environments. 

There is also evidence that hints toward a binary origin for repeating FRBs. 
The detected periodic activity from FRB~180916.J0158+65 \cite{chime16d} indicates a possible orbital motion of the FRB progenitor with a windy companion star \cite{myc+18}.
But pulsar binaries in the Galaxy often have a velocity of the order of 10-100~km~s$^{-1}$ and screen distance of light-seconds. 
Such a combination should produce variable large-curvature scintillation arcs, which is inconsistent with our current measurement.
Nevertheless, the discussions outlined above are solely based on the assumption of an isotropic screen or aligned V$_{\rm{eff}}$ vector. 
In reality, the scattering screen could be more complex than we assumed.

\subsubsection{Scintillation in MW}

Conversely, let us assume that the scintillation arc is caused by IISM in the MW  ($s \rightarrow$ 1), Eq.~\ref{arc_curature} can be simplified as:
\begin{equation}
\begin{split}
\eta \approx \frac{c d_{\rm{sc,MW}}}{2 \nu^{2}} \frac{1}{[(\boldsymbol{V_{\rm E}} - \boldsymbol{V_{\rm{sc}}}) \cos\boldsymbol{\psi}]^{2}}.
%\,\,\, \mathrm{s} \sim 1.
\label{eq:arc_definition_mw}
\end{split}
\end{equation}
According to the NE2001 electron density model of the Galaxy \cite{cl03}, there are two significant peaks in ionized medium density $C_{\rm n}^{2}$ at d$_{\rm{sc,MW}}$ $\approx$ 1.26 and 2.9 kpc, respectively.
Another possibility is that the scatter screen is from the local bubble located at a typical distance of 0.3~kpc from the Earth.
These are the three most probable locations for the scattering screen.

\begin{figure}[H]
    \centering
    \includegraphics[scale=0.4]{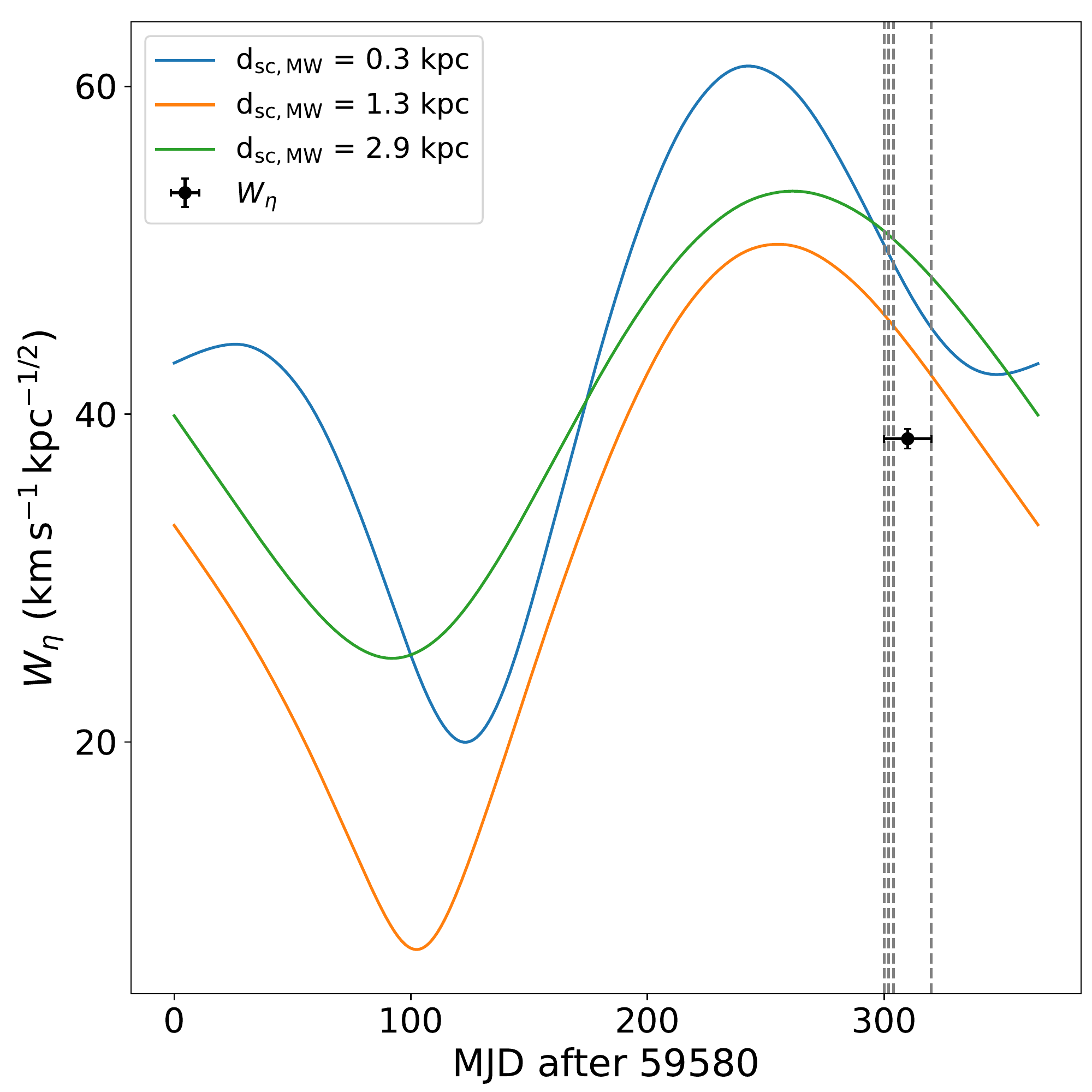}
    \caption{
    The possible annual variations in $W_{\eta}$ derived from scintillation arc curvature for a series  of screen distances from the earth, predicted with an assumption of scintillation occurring in MW and a stationary and isotropic screen. 
    The grey dashed lines represent the epochs with arc detection. 
    The black symbol represents the measured effective scintillation velocity $W_{\eta}$. The difference between the black point and the models could be caused by misalignment of V$_{\rm{eff}}$ and an anisotropic screen, or from additional velocity in the screen.}
    \label{fig:host_mw}
\end{figure}

Considering MW rotation speed as the intrinsic velocity of the screen,
we calculate the differential scaled velocity $W_{\eta}$ (Figure \ref{fig:host_mw}) between the scattering screen and the Sun assuming different screen distances with \textit{scintools}\footnote{\url{https://github.com/danielreardon/scintools}} \cite{rch+19, rcb+20}.
To simplify, we assume a scattering screen has no other motion except the intrinsic velocity from MW rotation and we further simplify the problem by taking $\cos(\psi)=1$, indicating $\boldsymbol{V_{\rm eff}}$ is aligned with the screen, or the screen is isotropic.
The result shows that Earth motion is the dominating factor, leading to a similar trend for all three assumed distances.

%The arcs of FRB~20220912A are detected within 20 days, 
%on which the predicted $W_{\eta}$ from arc curvature at a series of scattering screen distances are almost indistinguishable (see Figure~\ref{fig:host_mw}).
%The fact there are multiple solutions is coincidental.  
%The screen can be nearby with only the Earth's motion or distant, where the MW rotation compensates for the distance.

The DM$_{\rm{IGM}}$ contribution from the intergalactic medium is around $\sim$50 pc cm$^{-3}$ \cite{mpm+20} using the Redshift of 0.0771 \cite{rcc+22}. 
The MW contributed DM$_{\rm{MW}}$ is about 120 pc cm$^{-3}$ \cite{cl03, ymw16}. 
Thus, the combination of the host galaxy DM$_{\rm{hg}}$ and the Milky Way halo DM$_{\rm{halo}}$ is $\sim$50 pc cm$^{-3}$.
The scintillation arc indicates there is only one dominant scattering screen towards FRB~20220912A. %scintillation.
We think a Milky-way-origin of the screen is the more likely explanation. 
The detection of annual variation in scintillation timescale or arc curvature in future observations would be a smoking gun to test our Galactic screen interpretation. 
The maximum and minimum value of arc curvature $\eta$ would be observed in nearly April and September during a year, respectively.

\section{CONCLUSIONS}

During an active episode of FRB~20220912A with FAST, we clearly detected scintillation arcs for the first time from an FRB.
We find no significant variations in arc properties such as curvature and asymmetry distribution in power from observations spanning about 20 days, suggesting that the dominant screen and effective velocity remained stable over this time.
This could be explained by the RISS timescale $\tau_{\rm{r}}$ ($\geq$ 20 days \cite{ric90}), the timescale for feature movement across the arcs.

The scintillation strength favors a screen in MW.
We investigated these two scenarios by assuming various possible distances and velocities of the scattering screen, and an isotropic screen, i.e., cos{$\psi$}=1, 
and then favor a MW origin afterward.
However, the realities of FRB progenitor could differ from what we have assumed, indicating that our discussions are bounded. Thus the host galaxy origin cannot be ruled out.
If an MW screen causes the scintillation arc, we expect to observe an annual variation in the arc curvature or in the scintillation timescales.
Additionally, we could also infer the location of the dominant screen either from the annual arc/timescale variation \cite{mac+23, lmv+23} or from VLBI \cite{mlk+21}. 
 
The variations in scintillation timescale and arc have been a tool for studying the transverse orbital motion of pulsars \cite{lyn84, rcb+20}.
The first detection of a scintillation arc from an FRB demonstrates the possibility of studying the putative FRB orbital motion if we could find an FRB of which IISM causes scintillation in its host galaxy.

%On the other hand, the detected echoes from pulsars \cite{lpg01} suggest that epochs may be also detected from FRBs, which would worsen the spin period search process and be related to DM variations. 
%The scintillation arc-let \cite{hsa+05} can directly link the time offset between the main pulse and the echo pulse using the high-frequency resolution FRB data set.
%This could lead to discovering a spin period of an FRB.

%To conclude, scintillation arcs open a new window in the FRB field. Long-term scintillation arc monitoring could serve as the key to searching for possible FRB orbital motion.

%%%%%%%%%%%%%%%%%%%%%%%%%%%%%%%%%%%%%%%%%%%%%%%%%%%%%%%
%%% Acknowledgements. ??§Ý
%%%%%%%%%%%%%%%%%%%%%%%%%%%%%%%%%%%%%%%%%%%%%%%%%%%%%%%
\Acknowledgements{We thank the anonymous referees for the
valuable suggestions that improved this paper. 
This work made use of data from the FAST. 
FAST is a Chinese national mega-science facility, built and operated by the National Astronomical Observatories, Chinese Academy of Sciences. 
This work is supported by the National SKA Program of China No. 2020SKA0120200, 2020SKA0120100, % SKA
CAS Project for Young Scientists in Basic Research YSBR-063,
the National Nature Science Foundation grant No. 12041303, 11988101, 11833009, 11873067, 12041304, 12203045, and  
the National Key R$\&$D Program of China No. 2017YFA0402600, 2021YFA0718500, 2017YFA0402602, 2022YFC2205203, 
%CRAFTS
the CAS-MPG LEGACY project, the Max-Planck Partner Group, the Key Research Project of Zhejiang Lab no. 2021PE0AC0 and also the Western Light Youth Project of Chinese Academy of Sciences. 
}

%%%%%%%%%%%%%%%%%%%%%%%%%%%%%%%%%%%%%%%%%%%%%%%%%%%%%%%
%%% Conflict of interest. ????????????
%%%%%%%%%%%%%%%%%%%%%%%%%%%%%%%%%%%%%%%%%%%%%%%%%%%%%%%
\InterestConflict{The authors declare that they have no conflict of interest.}

%%%%%%%%%%%%%%%%%%%%%%%%%%%%%%%%%%%%%%%%%%%%%%%%%%%%%%%
%%% Appendix sections. ??????, ????
%%%%%%%%%%%%%%%%%%%%%%%%%%%%%%%%%%%%%%%%%%%%%%%%%%%%%%%

\end{multicols}

\newpage

\begin{appendix}
%\section{Name}

%\end{appendix}

%\begin{appendices}
%\section{Appendix}
%\end{appendices}
%\appendix

%\appendix

\renewcommand{\thesection}{Appendix}

\section{Arc Curvature Determination}
We here display the arc strength as a function of arc curvature and the scintillation arc region, both derived from Hough transform \cite{bot+16}. 

\begin{figure*}
    \centering    
    \includegraphics[width=0.99\linewidth]{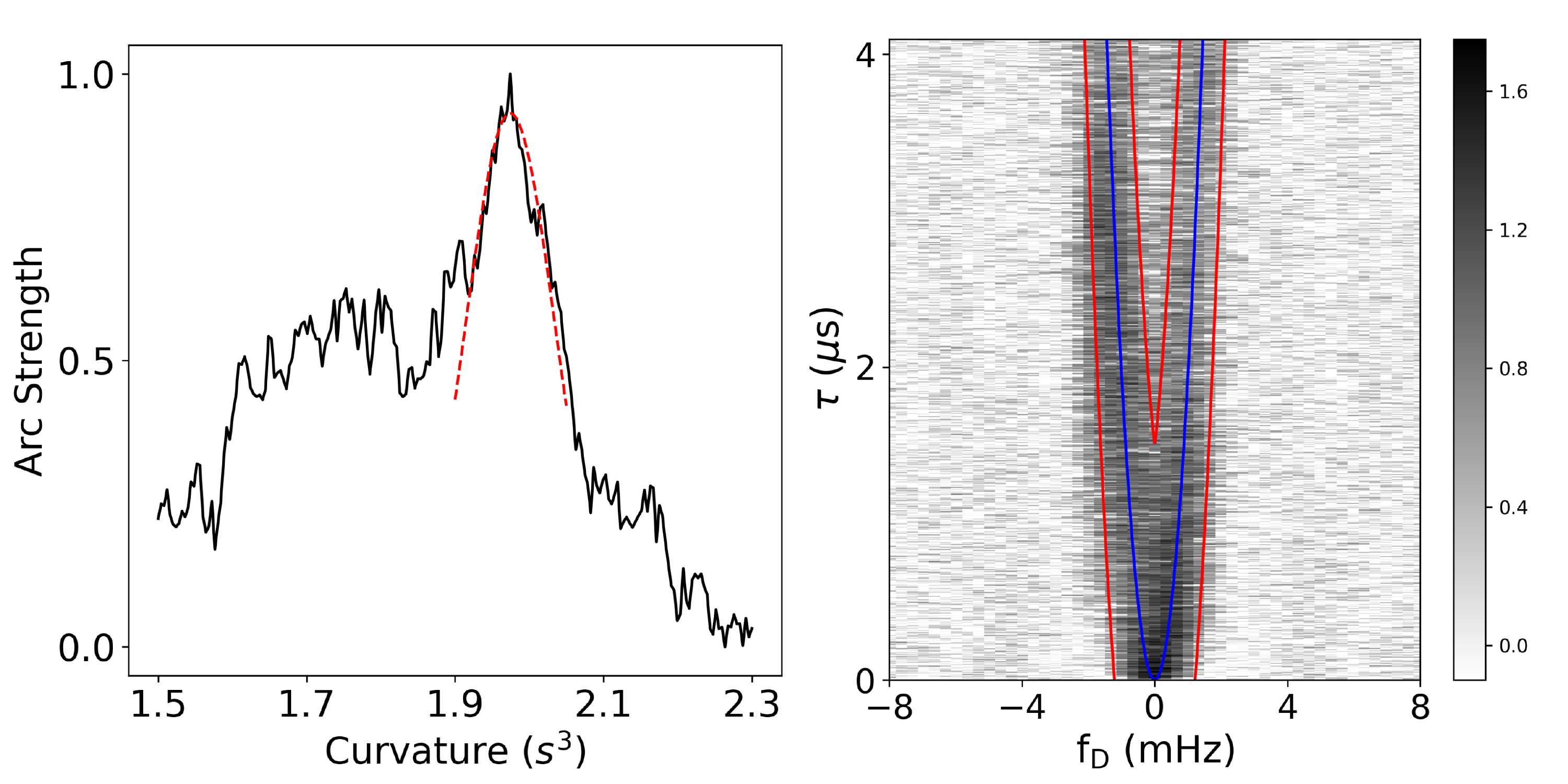} \\
    \caption{Arc curvature calculation using the Hough transform of FRB~20220912A on MJD~59882 with FAST. Left: arc strength as a function of curvature in black. The Gaussian curve in red is used to determine the arc curvature and its uncertainty (sigma of Gaussian). Right: the blue curve shows the parabola with the best curvature and the red curves enclose the detected scintillation arc
    .}
    \label{fig:hough}
\end{figure*}

%\end{appendices}
\end{appendix}


\begin{thebibliography}{99}
\bibitem{sch68}
Scheuer, P. A.~G. 1968, Nature, 218, 920

\bibitem{ric69}
Rickett, B.~J. 1969, Nature, 221, 158

\bibitem{rcb84}
Rickett, B.~J., Coles, W.~A., \& Bourgois, G. 1984, A\&A, 134, 390

\bibitem{sie82}
Sieber, W. 1982, A\&A, 113, 311

\bibitem{ric90}
Rickett, B.~J. 1990, Ann. Rev. Astr. Ap., 28, 561

\bibitem{lvm+22}
{Liu}, Y., {Verbiest}, J. P.~W., {Main}, R.~A., {et~al.} 2022, \aap, 664, A116

\bibitem{bts+22}
{Bignall}, H.~E., {Tuntsov}, A.~V., {Stevens}, J., {et~al.} 2022, \mnras, 513, 2770

\bibitem{wagw95}
{Wagner}, S.~J., \& {Witzel}, A. 1995, \araa, 33, 163

\bibitem{rlg97}
{Rickett}, Barney J. and {Lyne}, Andrew G. and {Gupta}, Yashwant. 1997, \mnras, 287, 739
 
\bibitem{smc+01}
{Stinebring}, D.~R., {McLaughlin}, M.~A., {Cordes}, J.~M., {et~al.} 2001, ApJ, 549, L97

\bibitem{bmg+10}
{Brisken}, W.~F., {Macquart}, J.~P., {Gao}, J.~J., {et~al.} 2010, \apj, 708, 232

\bibitem{rcb+20}
{Reardon}, D.~J., {Coles}, W.~A., {Bailes}, M., {et~al.} 2020, \apj, 904, 104

\bibitem{yzm+21}
{Yao}, J., {Zhu}, W., {Manchester}, R. N., {et~ al.} 2021, Nature Astronomy, 5, 788

\bibitem{wks+08}
{Walker}, M.~A., {Koopmans}, L.~V.~E., {Stinebring}, D.~R., \& {van Straten}, W. 2008, \mnras, 388, 1214

\bibitem{zty+23} 
{Zhang}, D., {Tao}, Z., {Yuan}, M., et al.\ 2023, Science China Physics, Mechanics, and Astronomy, 66, 299511

\bibitem{mpj+23}
{Main}, R.~A., {Parthasarathy}, A., {Johnston}, S., {et~al.} 2023, \mnras, 518, 1086
  
\bibitem{srm+22}
{Stinebring}, D.~R., {Rickett}, B.~J., {Minter}, A.~H., {et~al.} 2022, \apj, 941, 34

\bibitem{wvm+22}
{Wu}, Z., {Verbiest}, J. P.~W., {Main}, R.~A., {et~al.} 2022, \aap, 663, A116

\bibitem{lbm+07}
{Lorimer}, D.~R., {Bailes}, M., {McLaughlin}, M.~A., {Narkevic}, D.~J., \& {Crawford}, F. 2007, Science, 318, 777

\bibitem{mls+15}
{Masui}, K., {Lin}, H.-H., {Sievers}, J., {et~al.} 2015, \nat, 528, 523

\bibitem{rsb+16}
{Ravi}, V., {Shannon}, R.~M., {Bailes}, M., {et~al.} 2016, Science, 354, 1249

\bibitem{cordes+19} 
{Cordes}, J.~M. \& {Chatterjee}, S.\ 2019, \araa, 57, 417


\bibitem{cr98}
{Cordes}, J.~M., \& {Rickett}, B.~J. 1998, ApJ, 507, 846

\bibitem{wcv+23}
{Wu}, Z., {Coles}, W.~A., {Verbiest}, J. P.~W., {et~al.} 2023, \mnras, 520, 5536

\bibitem{occ+22}
{Ocker}, S.~K., {Cordes}, J.~M., {Chatterjee}, S., {et~al.} 2022, \apj, 931, 87

\bibitem{mbm+23} 
{Main}, R.~A., {Bethapudi}, S., {Marthi}, V.~R., {et~al.} 2023, \mnras, 522, L36


\bibitem{mc+22}
{McKinven}, R., \& {Chime/Frb Collaboration}. 2022, The Astronomer's Telegram, 15679, 1
  
\bibitem{ydn+22}
{Yu}, Z., {Deng}, F., {Niu}, C., {et~al.} 2022, The Astronomer's Telegram, 15758, 1

\bibitem{znf+22}
{Zhang}, Y., {Niu}, J., {Feng}, Y., {et~al.} 2022, The Astronomer's Telegram, 15733, 1

\bibitem{jyg+19}
{Jiang}, P., {Yue}, Y., {Gan}, H., {et~al.} 2019, Science China Physics, Mechanics, and Astronomy, 62, 959502

\bibitem{jth+20}
{Jiang}, P., {Tang}, N.-Y., {Hou}, L.-G., {et~al.} 2020, Research in Astronomy and Astrophysics, 20, 064

\bibitem{qys+20}
{Qian}, L., {Yao}, R., {Sun}, J., {et~al.} 2020, The Innovation, 1, 100053

\bibitem{rav22}
{Ravi}, V. 2022, The Astronomer's Telegram, 15716, 1

\bibitem{nzz+22}
{Niu}, J.-R., {Zhu}, W.-W., {Zhang}, B., {et~al.} 2022, Research in Astronomy and Astrophysics, 22, 124004

\bibitem{ran11}
{Ransom}, S. 2011, Astrophysics source code library, ascl 2, 3, 4

\bibitem{zlz+23} 
{Zhang}, Y.-K., {Li}, D., {Zhang}, B., et al.\ 2023, \apj, 955, 142


\bibitem{mhm+22}
{Main}, R.~A., {Hilmarsson}, G.~H., {Marthi}, V.~R., {et~al.}
  2022b, \mnras, 509, 3172

\bibitem{zhz+22}
{Zhou}, D.~J., {Han}, J.~L., {Zhang}, B., {et~al.} 2022, Research in Astronomy and Astrophysics, 22, 124001 

\bibitem{rch+19}
{Reardon}, D.~J., {Coles}, W.~A., {Hobbs}, G., {et~al.} 2019, \mnras, 485,
  4389

\bibitem{wms+04}
{Walker}, M.~A., {Melrose}, D.~B., {Stinebring}, D.~R., \& {Zhang}, C.~M. 2004, \mnras, 354, 43

\bibitem{crs+06}
{Cordes}, J.~M., {Rickett}, B.~J., {Stinebring}, D.~R., \& {Coles}, W.~A. 2006, \apj, 637, 346

\bibitem{rcc+22}
{Ravi}, V., {Catha}, M., {Chen}, G., et al.\ 2023, \apjl, 949, L3


\bibitem{crg+10}
{Coles}, W.~A., {Rickett}, B.~J., {Gao}, J.~J., {Hobbs}, G., \& {Verbiest},
  J.~P.~W. 2010, ApJ, 717, 1206

\bibitem{bot+16}
{Bhat}, N.~D.~R., {Ord}, S.~M., {Tremblay}, S.~E., {McSweeney}, S.~J., \&
  {Tingay}, S.~J. 2016, \apj, 818, 86

%\bibitem{zb20} {Zhang}, B.\ 2020, \nat, 587, 45

\bibitem{brb+20} 
{Bochenek}, C.~D., {Ravi}, V., {Belov}, K.~V., {et al.}\ 2020, \nat, 587, 59

\bibitem{msh+18} 
{Michilli}, D., {Seymour}, A., {Hessels}, J.~W.~T., {et al.}\ 2018, \nat, 553, 182

\bibitem{ll94} 
{Lyne}, A.~G. \& {Lorimer}, D.~R.\ 1994, \nat, 369, 127

\bibitem{alp75} 
{Anderson}, B., {Lyne}, A.~G., \& {Peckham}, R.~J.\ 1975, \nat, 258, 215

\bibitem{lpg01} 
{Lyne}, A.~G., {Pritchard}, R.~S., \& {Graham-Smith}, F.\ 2001, \mnras, 321, 67

\bibitem{chime16d} 
{Chime/Frb Collaboration}, {Amiri}, M., {Andersen}, B.~C., {et al.}\ 2020, \nat, 582, 351

\bibitem{myc+18} 
{Main}, R., {Yang}, I.-S., {Chan}, V., {et al.}\ 2018, \nat, 557, 522

\bibitem{cl03}
{Cordes}, J.~M. and {Lazio}, T.~J.~W 2003, arXiv, astro-ph/0301598, astro-ph/0301598

\bibitem{mpm+20} 
{Macquart}, J.-P., {Prochaska}, J.~X., {McQuinn}, M., {et al.}\ 2020, \nat, 581, 391

\bibitem{ymw16} 
{Yao}, J.~M., {Manchester}, R.~N., \& {Wang}, N.\ 2017, \apj, 835, 29

\bibitem{mac+23} 
{Main}, R.~A., {Antoniadis}, J., {Chen}, S., et al.\ 2023, \mnras, 525, 1079


\bibitem{lmv+23}
{Liu}, Y., {Main}, R.~A., {Verbiest}, J.~P.~W., et al.\ 2023, Science China Physics, Mechanics, and Astronomy, 66, 119512


\bibitem{mlk+21} 
{Main}, R., {Lin}, R., {van Kerkwijk}, M.~H., et al.\ 2021, \apj, 915, 65

\bibitem{lyn84}
{Lyne}, A.~G. 1984, Nature, 310, 300


\end{thebibliography}
\end{document}